**Evidence of photo-induced dynamic competition of metallic and insulating phase in a layered manganite**

Yuelin Li [1*], Donald A. Walko[1], Qing'an Li[2], Yaohua Liu [2,3], Stephan Rosenkranz[2], Hong Zheng [2], and J. F Mitchell[2]

*ylli@aps.anl.gov

[1]Advanced Photon Source, Argonne National Laboratory, Argonne, IL 60439, USA

[2]Materials Science Division, Argonne National Laboratory, Argonne, IL 60439, USA

[3]Quantum Condensed Matter Division, Oak Ridge National Laboratory, TN 37831, USA

We show evidence that the competition between the antiferromagetic metallic phase and the charge- and orbital-ordered insulating phase at the reentrant phase boundary of a layered manganite, $LaSr_2Mn_2O_7$, can be manipulated using ultrafast optical excitation. The time-dependent evolution of the Jahn-Teller superlattice reflection, which indicates the formation of the charge and orbital order, was measured at different laser fluences. The laser-induced enhancement and reduction the Jahn-Teller reflection intensity shows a reversal of sign between earlier (~10 ns) and later (~150 ns) time delays during the relaxation after photo excitation. This effect is consistent with a scenario whereby the laser excitation modulates the local competition between the metallic and the insulating phases.

*ylli@aps.anl.gov

The essence of strongly correlated materials stems from competition among symmetry-broken quantum states, such as high-temperature superconductivity, and charge-, spin- and orbital-ordered phases[1]. Among many strongly correlated materials, colossal magnetoresistive (CMR) manganites are a prototypical system with complex couplings among charge, orbital, spin, and lattice orders. It is generally believed that local lattice distortions of the $MnO_6$ octahedra accompany anti-ferromagnetic, charge- and orbital- ordered (COO) phases on the one hand, and the mixed-valence character promotes metallic ferromagnetic double exchange on the other hand. The transport properties are determined by the mutual competition of these phases. They have been widely studied due to the intriguing physics and technological potential for magnetic memory, logic, etc., arising from their half-metallicity controllable via both magnetic field and light. By directly or indirectly modifying different degrees of freedom, a plethora of experiments have demonstrated the feasibility of using ultrafast laser pulses to create and probe novel transient states that are inaccessible via adiabatic or static thermal methods [2–4].

Manganites come in several families of structures. The reduced dimensionality in bilayer manganites ($La_{2-2x}Sr_{1+2x}Mn_2O_7$) [5,6] amplifies the effect of entropy change in the temperature regime near the phase transitions and offers a unique opportunity to study the ordering dynamics. For x close to 0.5, a COO insulator phase is sandwiched between a low-temperature antiferromagnetic metallic (AFMM) phase and a high-temperature (> 220 K) paramagnetic metallic (PMM) phase (Fig. 1 (a)). The AFMM phase transforms into the COO phase when the transition temperature is approached from below, representing a spin-charge order critical phase point observed only in a few manganite compounds[7–10]. At x~0.5, the COO is of checkerboard form characterized by a $4 \times 4$ superlattice, whereas at x~0.6, the COO has been shown by x-ray resonant scattering to adopt a bistripe-order of a $5 \times 5$ superlattice [11–13].

There are many studies, under adiabatic or static conditions, on the complex ground states in the AFMM phase[14,15], the formation of the COO[11–13] the COO-associated Jahn-Teller (JT) distortion[11,12], and the correlation between the magnetization and the transport properties [16,17].

In this paper, we report evidence of non-thermal dynamic control of the metallic and the COO insulator phases in the x=0.505 compound using ultrafast laser pulses by measuring the COO-associated JT distortion dynamics. By varying the laser fluence we were able to either enhance or reduce the transient JT x-ray reflection, which is consistent with photo-induced mutual phase conversion between the metallic and insulator phases. This phase conversion has been studied in $Pr_{0.55}(Ca_{0.75}Sr_{0.25})_{0.45}MnO_3$ thin film samples [9,10] via monitoring the transport and optical spectroscopy. The current pump-probe experiment provides direct structural evidence of the competition between the photoexcited metallic and COO insulator phases.

$LaSr_2Mn_2O_7$ has a tetragonal structure with a=b=3.87 Å and c=20.1 Å. The sample, 2 mm by 2.5 mm by 0.7 mm in size, was grown using the floating zone technique[18]. A phase diagram around x=0.5 is shown in Fig. 1 (a). The sample surface is naturally cleaved. The magnetization was characterized by a SQUID magnetometer. The x-ray measurements were conducted at the 7-ID beam line at the Advanced Photon Source (APS)[19].

The static temperature dependence of the COO was measured at the (0.75 1.25 6) JT superlattice reflection that is due to the localization of the $Mn^{3+}$ $e_g$ electrons[11,12]. The temperature dependence of the JT reflection intensity is compared with the magnetization in Fig. 1(b, c). The comparison is very similar to that between the transport and the magnetization (Fig. 2 in [17]), but with a transitional temperature region (labelled as TR throughout this paper) between 100 K and 135 K, where the JT reflection gradually increases to the maximum. The JT superlattice reflection peak position was observed to have a negligible dependence on the temperature.

For the pump probe measurement, the sample was cooled down to 16 K and then raised to and kept at 49 K in a Displex cryogenic stage, putting the system on the warming loop of Fig. 1 (c). Laser pulses with duration less than 100 fs at 800 nm (1.55 eV) were delivered to the sample at normal incidence through a fused silica window at a repetition rate of 5 kHz, with a spot size of about 2.5 mm by 2 mm (full width at half maximum). The laser penetration depth was estimated to be about ~50 nm using the optical data for a composition with close doping level from ref.[20]. To match the x-ray (12 keV) and laser penetration depths, the x-ray incidence angle was set at 0.22 degree[21]. Note that the scattering geometry is asymmetric and the x-ray exit angle is 19°. An example of the diffraction peak is depicted in Fig. 1 (d), showing a broad, but well-defined peak with no spurious contamination.

The experiment was performed in the 24-bunch mode of the APS with a bunch-to-bunch spacing of 153 ns. An avalanche photodiode gated at the laser repetition rate of 5 kHz with 4 gates was used to record the diffraction data. For the "early" channel, the gate was set to observe the bunch that coincided with the laser pump pulse. The "late" channel was further delayed by 153 ns to gate the next bunch in the storage ring. Two additional gates were set at 3.682 $\mu$s before the early and late channels, i.e., one round trip of the storage ring prior, to measure the "no-laser" signal from the same photon bunches. The incoming beam intensity was also monitored for normalization of the x-ray counts. The x-ray pulse duration is about 100 ps. Further details of the diffraction measurements are provided in the Supplemental Materials [21].

The pump-probe measurement was performed by changing the time delay of the laser pump relative to the x-ray probe at several laser fluences. The "no-laser" JT diffraction signal at (0.75 1.25 6) was found to be nontrivially dependent on the laser fluence, shown in Fig. 2. This is the intensity $I$ of the JT reflection at 3.68 μs before the arrival of the laser pulse or, equivalently, 196 μs after the previous laser pulse. This "no-laser" JT reflection was stable during the measurement at each laser fluence (lasts about 25 min), as shown in Fig. 2 by the error bars.

This fluence dependence of the JT reflection has features somewhat similar to its temperature dependence in Fig. 1 (c), indicating a slow recovery of the sample from the laser excitation. The change of the JT intensity is likely a convolution of changes in the JT structure factor and the relative population of the COO phase representing an effective "initial state" for the pump probe measurement.

The transient difference signal, $\Delta I(t)=I(t)-I(t-3.6\ \mu s))$, i.e., the signal with the "no-laser" signal (Fig. 2) subtracted, is presented in Fig. 3 at several laser fluencies as a function of delay between the laser and the x-ray. Positive or negative signals indicate an enhancement or reduction of the JT reflection, respectively. $\Delta I$ depends on both the delay and the laser fluence, which is summarized in Fig. 4, showing the average difference signal in the time windows of 6-10 ns (early) and 159-163 ns (late) after the laser excitation. For all fluences, $\Delta I$ before laser pulse is essentially zero. Note that the signal level of a few percent can be compared with the excitation efficiency estimated from the laser fluence, i.e., about 6% Mn ions in the top 50 nm probed layer can be excited at 1 mJ/cm$^2$ (40 photons per nm$^2$) assuming 100% absorption [21].

There is a striking feature in Fig. 4: the early and the late $\Delta I$ have opposite signs. At low laser fluencies (<3.3 mJ/cm$^2$), the early $\Delta I$ is positive but the late $\Delta I$ is negative. At fluencies between 4.2 and 6.3 mJ/cm$^2$, a zero or negative early $\Delta I$ is observed, albeit accompanied by a positive late $\Delta I$. Clearly the laser excitation can enhance or reduce the JT reflection instantaneously, but the relaxation that follows can reverse this trend in 150 ns.

To understand the result in Fig. 4, it is critical to characterize the pseudo ground state as a function of the laser fluence, as shown in Fig. 2. To do this, we assume that after 0.2 ms, residual heating effect dominates the sample bulk. Thus we apply a two-temperature, 1-D thermal heat diffusion model for the actual sample thickness of about 0.7 mm, for which the detail is given in SI [21].

The simulation is benchmarked by the 7.74 mJ/cm$^2$ case. We note in the data shown in Fig. 3, the JT intensity has an instantaneous drop (<100 ps), of about 15%, followed by a slow decrease over more than 100 ns. The instantaneous reduction can be assigned as non-thermal, photon excitation of the COO states that destroys the JT distortion. The relaxation of these excited states, however, leads to slow heating of the sample partially beyond the COO-PMM transition temperature ($T_{CP}$) thus further decrease of the JT intensity. Therefore, the sample temperature before the laser pulse must be below $T_{CP}$ at 220 K. The later time (160 ns) temperature, however, must exceeds $T_{CP}$ for the JT intensity to further decrease. Using the thermal simulation, we can determine a total heating efficiency of 36%. The spatial temporal evolution of the bulk temperature is shown in Fig. 5 (a) using a CW simulation (For detailed explanation, see the supplementary information [21]), where the sample temperature reaches a equilibrium at 180 K within 2 seconds. The corresponding temperature evolution due to pulsed laser excitation (assuming all absorbed laser energy convert to heat immediately) is shown in Fig. 5 (b). The temperature as a function of the laser fluence is shown in Fig. 5 (c), where the temperature for the first 100 nm of the sample before, at 8 ns (early) and 160 ns (late) after the laser exposure are given. We can see that at a fluence above 4 mJ/cm$^2$, if all the energy is immediately converted to heat, the temperature would be higher than $T_{CP}$, leading 100% reduction of the JT reflection, contradicting the experiment observation. This confirms that the instantaneous excitation is non-thermal. For the 7.7 mJ/cm$^2$ case the late temperature is slightly above $T_{CP.}$ These simulations also indicate that after the initial excitation, the sample has a tendency to heat up probably due to the relaxation of the excited states via phonon emission.

The thermal simulation confirms that the state before the each laser exposure is indeed thermally stabilized, which allows a direct comparison of the initial state to the static mapping of the JT intensity to a temperature dependence, as shown in Fig. 2 (b). It shows a clear trend of the initial dominating phase shifting from AMM to COO phases as laser fluence increases. Interestingly, the highest JT reflection reaches the maximum in the TR region and is not directly correlated with static peak position, but are within a factor of two of these in the COO phase. Due to the stable signal we observed at each laser fluence, this is unlikely due to spurious experiment errors. However, the underlying physics needs further study. We also noticed the JT signal at laser fluence lower than 1 mJ/cm$^2$, which we attribute to the laser excitation of a lower temperature phase with JT distortion that is stable at temperature lower than 75 K as reported in [11,15].

With the initial condition determined, and noting that non-thermal photo excitation has been observed to convert metal into insulator [10,22,23] and vice versa in other manganites and oxides [9,24,25], we propose that the observed phenomena in Fig. 4 as evidence of photon driven phase competition between the metallic and the insulating phases. We propose that the quasi-equilibrated "initial state" before the arrival of the laser pulse is phase-separated, a common scene in many strongly correlated oxides[1], with coexisting AFMM and COO clusters up to the AMM-COO phase transition at about $T_{AC}$=136 K.

In this case, both AFMM-to-COO and the COO-to-metal phase conversions occur upon laser excitation at least up to $T_{AC}$ as has been reported in previous experiments [9,10]. The former conversion will enhance and the latter will reduce the intensity of the JT reflection. Assuming that the excitation rates are about the same for both conversions, the net change in the JT reflection immediately after the laser pulse is thus determined by the fraction of each phase in the "initial state." Qualitatively, if the initial state is dominated by the AFMM phase, more COO clusters will be generated than melted, leading to an immediate enhancement of the JT reflection. When the COO dominates, however, more COO clusters will be melted than generated, thus a net immediate reduction of the JT intensity. At the same time, the initial fractional contribution is dependent on the equilibrium temperature of the sample and due to the surrounding structure and spin "pressure", as coined in refs. [9,10]. It is expected that the newly generated non-dominant phase clusters, i.e, COO clusters at a low laser fluence and metallic clusters at a high laser fluence, decay faster than the newly generated dominant phase clusters, i.e., AFMM clusters at a low fluence and COO clusters at a high fluence. Our experiments are not able to determine the final form of the local electronic state immediately after melting of the COO clusters. At a low laser fluence, it is likely spin-ordered due to the low temperature dominant AFMM phase or will become spin-ordered soon due to surrounding "spin pressure"[9]. At a high laser fluence, these metallic clusters can be PMM.

Based on this argument, we can now qualitatively interpret the observations in Fig. 4. At low laser fluence the JT reflection sees an immediate enhancement after the laser excitation because more COO insulator clusters are generated than destroyed as the initial state is dominated by the AFMM phase. The difference signal turns negative later when these laser-generated COO clusters decay while the laser-generated metallic clusters stay. At high fluence, the behaviors of the newly generated COO and metallic phases reverse, and the opposite trend is seen. It is expected that the trend reverses at an initial state where sample reaches AFMM-COO transition at $T_{AC}$. Our data also indicate a relaxation time of the laser-generated non-dominant phase of about 100-200 ns.

Non-thermal mechanisms modifying both local spin order and the COO phase due to photo excitation have been widely discussed in the literature though the details are still under discussion. The complication here is the competition between the COO insulating and the spin-ordered metallic phases. Melting of the COO phase[9,24–26] is likely due to charge transfer of $t_{2g}$ electrons into the $e_g$ orbit thru either an intra-site [27] or an inter-site [28,29] d-d transition, leading to an increase of the electron itinerancy which couples to the phonon and releases the JT distortion. Evidence further shows that it is also possible to induce magnetization in manganites via photo-excitation [26,30]; thus in the low fluence AFMM-dominated phase, the surrounding spin "pressure" may coalesce the metallic clusters into the spin-ordered phase. At the high fluence COO-dominated phase, melting routes to either PMM or AFMM phases are possible. The formation of the COO phase is accompanied by melting of the AFMM spin order, i.e., ferromagnetic order in the a-b plane. It is suggested that a non-thermal melting of the spin order [10] can be accomplished via photo induced spin-flip for the $e_g$ electrons against the Hund coupling energy [31,32]. This spin flip may lead, within the ferromagnetic plane, to a local spin correlation favored by the COO and the JT distortion.

The physics picture of photon manipulation of the competition between the metallic and the insulating phases[9,10] qualitatively explained the dynamic behavior, but cannot fully explain the higher JT intensity in the TR region. Although bilayer manganite presents an array of exotic properties, there is surprisingly very limited study of dynamic properties and this work is, to our knowledge, the first study of such dynamics of this material. Clearly, further effort is needed to understand the result. It will be very illuminating to selectively excite the lattice at the stretched mode at 17 μm [33] in either the metallic or the COO-dominated phase to clarify the role of lattice distortion in facilitating the complex coupling and competition in the system [2,4,34].

In summary, we have shown evidence of the dynamic competition between spin and charge order in the bilayer manganite $LaSr_2Mn_2O_7$. Our result is qualitatively consistent with a scenario wherein non-thermal laser excitation leads to the transient modulation of the relative fractions of the metallic and the COO insulating phases in an effective phase-separated initial state. It also shows that the relative relaxation time is important in determining system dynamics. The experiment suggests the feasibility of controlling the competition between the two phases at a much faster time scale than previously observed[9,10].

Work performed at the ANL was supported by U.S. Department of Energy, Office of Science, Office of Basic Energy Sciences, under Contract No. DE-AC02-06CH11357. Work at QCMD, ORNL is supported by the Division of Scientific User Facilities of the Office of Basic Energy Sciences, US Department of Energy.

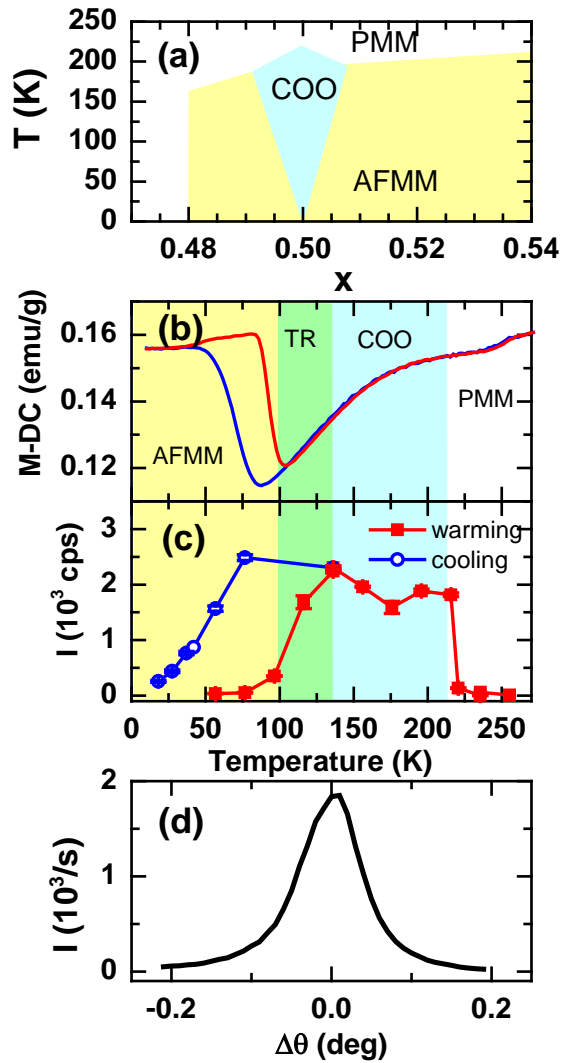

Figure 1. (a) Schematic phase diagram of the bilayer manganite at around x=0.5, after Fig. 1 in ref. [18], our sample is at x~0.505. Temperature dependences of (b) sample magnetization and (c) intensity of J-T reflection at (0.75, 1.25, 6). Blue and red curves indicate cooling and warming curves, respectively. The J-T reflection intensity is measured from the same detector channel as the one used in Fig. 2, thus the intensity can be directly compared. The warming and cooling data were collected with x-ray incidence angles at 10º and 0.22º, respectively. The warming curve is scaled to that at 0.22º at the overlapping temperature region. The dominating phases at the respective temperature are shaded with color, with a transitional region marked as 'TR', where the JT reflection continuous to increase. (d) shows the profile of the JT diffraction peak at 22 K at θ=0.22º x-ray incidence angle.

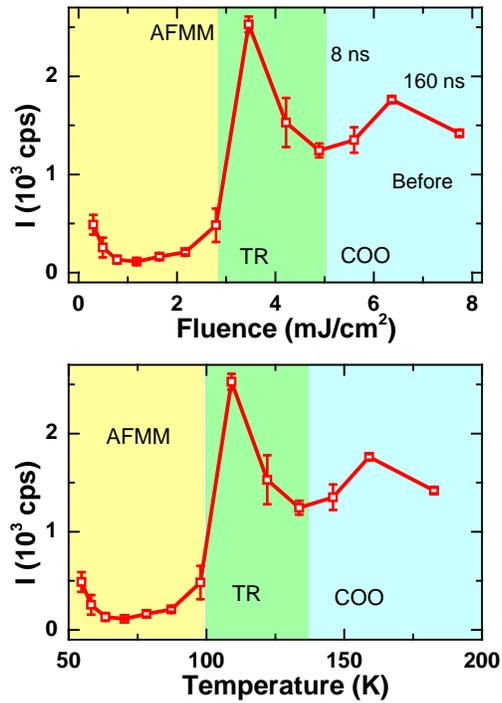

Figure 2 The JT reflection intensity of the average intensity (red) of the (0.75 1.25 6) JT reflection, taken 3.7 μs before the arrival of the laser pulse or, equivalently, 196 μs after the previous laser pulse (red) as (a) a function of the laser-fluence and (b) as a function of the calculated temperature. This is the "no-laser" signal referred in the text. The dominating phases are shaded with different colors.

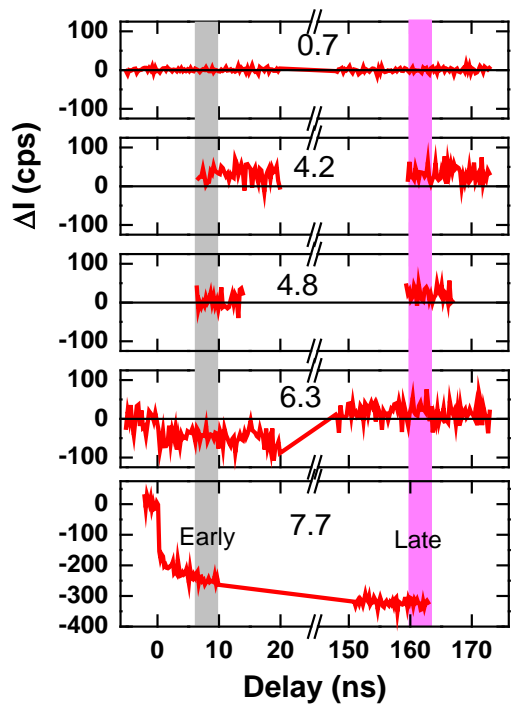

Figure 3 The difference signal ΔI at the JT reflection (0.75 1.25 6) as a function of delay between the X-ray and the laser pulses at various laser fluences (marked in mJ/cm$^2$ in each panel). The gaps in the 4.2 and 4.8 mJ/cm$^2$ data are due to beam losses. The "early" and "late" time windows are shaded. The x-ray beam is controlled and monitored by a set of beam position monitor feedbacks and the beam loss does not change the beam size, beam alignment, as well as the timing between the laser and x-ray. It did not otherwise affect the data collection.

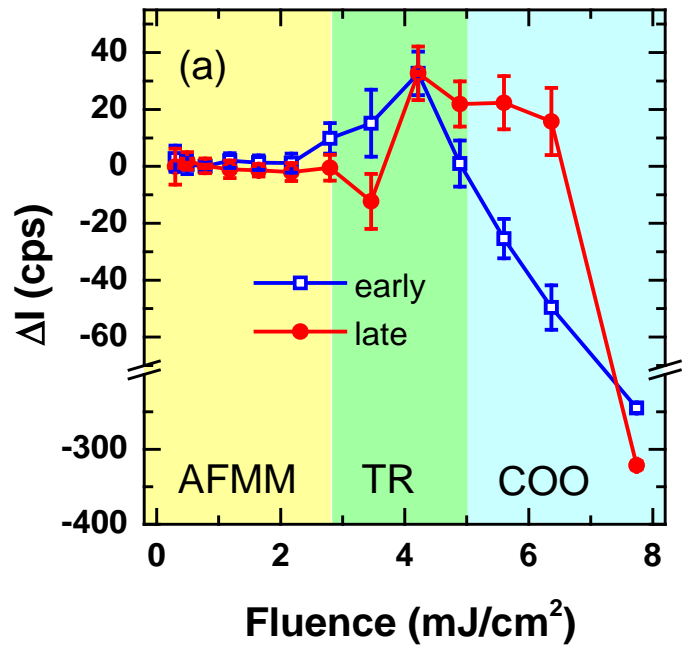

Figure 4 The difference intensity of the JT reflection at (0.75 1.25 6) averaged over two time windows as a function of the laser fluence, early: 5-10 ns after laser pulse; late: 158-163 ns after the laser pulse.

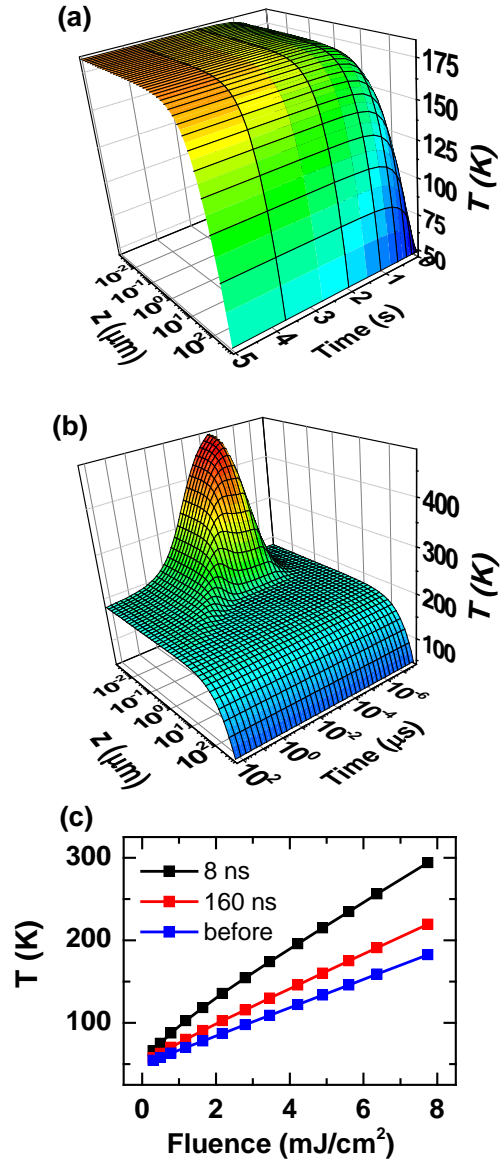

Figure 5. CW (Top) and pulsed (middle) thermal simulation of the sample temperature evolution at 7.7 mJ/cm$^2$. The temperature stabilizes within 2 s at 180 K. Laser incidence from the left. (Bottom) Temperature as a function of the laser fluence at different times after the laser pulse, assuming a total heating efficiency of 36% based on the benchmark using the data at 7.7 mJ/cm$^2$. For fluence larger than 4.5 mJ/cm$^2$, the early temperature far exceeds the COO-PMM transition temperature, excluding heating as the possible cause of the early signal.

**Supplemental materials**

**X-ray parameters and setup procedure**

The x rays were focused by Kirkpatrick-Baez mirrors; the spot size is estimated to be 50 µm by 50 µm on the sample. The beam divergence of 0.04° was dominated by the focusing optics. With the small incidence angle of 0.22°, the x-ray foot print overfilled the sample, so the laser spot size was set to illuminate the whole sample (~2 mm by 2.5 mm).

The x-ray beam was controlled and monitored by a beam-position monitor operating with feedback; thus, the beam losses that occurred during the measurements did not change the beam size or beam alignment, nor the timing between the laser and x-ray.

The temperature control was via a Displex cryostat (Advanced Research Systems). The sample was attached to the sample stage using silver paint. In all the delay scans, the recorded sample stage temperature variation was less than 0.2 K. With the temperature changing from 80 K to 30 K, the sample height changed less than 20 µm, mostly due to the temperature change of the sample holder; therefore position change due to laser irradiation is not expected. The compressor connected to the Displex, however, did cause the stage to vibrate horizontally at amplitude of about 50 µm or less at about 1 Hz. However, this movement is averaged out during the 10 to 20-s data collection time. Our temperature scan has demonstrated that there is no drift of the (0.75 1.25 6) peak (within our resolution) as the temperature is changed, indicating the thermal expansion of the sample is very small.

A six-circle Huber diffractometer was used in the experiment to capture the diffraction peaks. The Huber includes sample angles of pitch (Eta), roll (Chi), and yaw (Phi), and detector horizontal (Nu) and vertical (Del) angles, and a sample- table horizontal rotation (Mu). The diffractometer was controlled with SPEC (Certified Scientific Software) during the experiment using psi-circle geometry [1][S1]. We used the (0 0 6) and (1 1 6) Bragg peaks to determine the orientation matrix, from which it was straightforward to find the (0.75 1.25 6) superstructure peak. We operated in a SPEC mode in which the x-ray incidence angle can be selected; we set the x-ray incidence angle at 0.22° for all measurements in this experiment. The most important systematic error is due to the uncertainty in determining the zero incidence angle for the sample in the first step due to the small sample size and beam divergence. We estimate the error to be around 0.05° using the zero degree cut method, however, the terraces exists in the naturally cleaved sample surfaces may also add to this error. Together with the x-ray beam divergence, we are expecting larger uncertainties in the incidence angle. The configuration of the diffractormeter was maintained throughout each measurement once set and we do not expect any deviation from the set positions or angles for either the sample or the x-ray detector.

**Two temperature 1-D thermal simulation of the static fluence dependent initial state**

A two temperature heat diffusion model is described by the following equations

$$C_e(T_e)\frac{\partial T_e}{\partial t} = \frac{\partial}{\partial x}\left[K_e(T_e,T_l)\frac{\partial T_e}{\partial x}\right] - G(T_e)(T_e - T_l) + S(x,t), \tag{S1}$$

$$C_l(T_l)\frac{\partial T_l}{\partial t} = \frac{\partial}{\partial x}\left[K_l(T_e,T_l)\frac{\partial T_l}{\partial x}\right] + G(T_e)(T_e - T_l), \tag{S2}$$

Here $T_e$, $T_l$ are electron and phonon temperatures, $C_e$ and $C_l$ are the electron and phonon specific heat, $K_e$ and $K_l$ are the electron and phonon conductivity. $G$ is the coupling coefficient between electron and phonons. $S$ is the heat source term.

    The heat capacity as a function of temperature from Gu et al.,[2] for x=0.5 compound is used. For c-axis phonon heat conductivity as a function of temperature, we use the data measured by Matsukawa et al [3] for an x=0.6 compound. The c-axis electron heat conductivity is at least 2 orders of magnitude smaller [3] and is ignored. The a-b plane phonon conductivity is about 3 times bigger but as the side of the sample is only in contact with vacuum thus justifies the 1 D model. For the electron-phonon coupling we used the measured number G= 5e15 W/(m$^3$ K) from a perovskite compound La$_{0.7}$M$_{0.3}$MnO$_3$ by Averitt et al [4]. The boundary temperature is set at the sample mount temperature of 49 K.

    A laser pulse of 60 fs duration is used which deposit the energy in the sample with an absorption length of 50 nm. Two types of simulation are performed. The first one is used to determine the equilibrium temperature of the sample with a continuous laser flux equivalent to the average laser power (CW simulation), the second type uses a pulsed heat source emulating the laser to examine the temperature evolution for each heating cycle (pulsed simulation) taken the temperature profile of the CW simulation as the initial condition. Cross check between the two types of simulation gives excellent agreement.

To calculate the laser absorption, we use the complex refractive index at 800 nm of 1.59+1.29i is measured for x=0.34 composition, ref. [5].

**Estimate of the nonthermal laser excitation effect**

For direct photoexcitation, assuming an excitation event corresponds to absorption of one photon by one Mn ion, a pump fluence of 1 mJ cm$^{-2}$ corresponds to an excitation rate of 6% for a 50 nm surface layer. At 7 mJ/cm$^2$, the effect is 40%. This is consistent with the signal level we observe.